\documentclass[onecolumn,preprintnumbers,amsmath,amssymb]{revtex4}

\usepackage{graphicx}
\usepackage{dcolumn}
\usepackage{bm, amsmath, amssymb}

\usepackage[caption=false]{subfig}

\usepackage{latexsym,verbatim,color}
\usepackage{theorem}

\def\squareforqed{\hbox{\rlap{$\sqcap$}$\sqcup$}}
\def\qed{\ifmmode\squareforqed\else{\unskip\nobreak\hfil
\penalty50\hskip1em\null\nobreak\hfil\squareforqed
\parfillskip=0pt\finalhyphendemerits=0\endgraf}\fi}
\def\endenv{\ifmmode\;\else{\unskip\nobreak\hfil
\penalty50\hskip1em\null\nobreak\hfil\;
\parfillskip=0pt\finalhyphendemerits=0\endgraf}\fi}

\mathchardef\ordinarycolon\mathcode`\:
\mathcode`\:=\string"8000
\def\vcentcolon{\mathrel{\mathop\ordinarycolon}}
\begingroup \catcode`\:=\active
  \lowercase{\endgroup
  \let :\vcentcolon
  }

\newcommand{\nc}{\newcommand}
\nc{\rnc}{\renewcommand}
\nc{\beq}{\begin{equation}}
\nc{\eeq}{{\end{equation}}}
\nc{\beqa}{\begin{eqnarray}}
\nc{\eeqa}{\end{eqnarray}}
\nc{\lbar}[1]{\overline{#1}}
\nc{\bra}[1]{\langle#1|}
\nc{\ket}[1]{|#1\rangle}
\nc{\ketbra}[2]{|#1\rangle\!\langle#2|}
\nc{\braket}[2]{\langle#1|#2\rangle}
\nc{\proj}[1]{| #1\rangle\!\langle #1 |}
\nc{\avg}[1]{\langle#1\rangle}
\nc{\Rank}{\operatorname{Rank}}
\nc{\smfrac}[2]{\mbox{$\frac{#1}{#2}$}}
\nc{\Tr}{\operatorname{Tr}}
\nc{\tr}{\operatorname{Tr}}
\nc{\id}{\operatorname{id}}
\nc{\1}{\openone}
\nc{\ox}{\otimes}
\nc{\dg}{\dagger}
\nc{\dn}{\downarrow}
\nc{\cA}{{\cal A}}
\nc{\cB}{{\cal B}}
\nc{\cC}{{\cal C}}
\nc{\cD}{{\cal D}}
\nc{\cE}{{\cal E}}
\nc{\cF}{{\cal F}}
\nc{\cG}{{\cal G}}
\nc{\cH}{{\cal H}}
\nc{\cI}{{\cal I}}
\nc{\cJ}{{\cal J}}
\nc{\cK}{{\cal K}}
\nc{\cL}{{\cal L}}
\nc{\cM}{{\cal M}}
\nc{\cN}{{\cal N}}
\nc{\cO}{{\cal O}}
\nc{\cP}{{\cal P}}
\nc{\cQ}{{\cal Q}}
\nc{\cR}{{\cal R}}
\nc{\cS}{{\cal S}}
\nc{\cT}{{\cal T}}
\nc{\cX}{{\cal X}}
\nc{\cY}{{\cal Y}}
\nc{\cZ}{{\cal Z}}
\nc{\supp}{{\operatorname{supp}}}
\nc{\var}{\operatorname{var}}
\nc{\rar}{\rightarrow}
\nc{\lrar}{\longrightarrow}
\nc{\polylog}{\operatorname{polylog}}

\nc{\RR}{{{\mathbb R}}}
\nc{\CC}{{{\mathbb C}}}
\nc{\FF}{{{\mathbb F}}}
\nc{\NN}{{{\mathbb N}}}
\nc{\ZZ}{{{\mathbb Z}}}
\nc{\PP}{{{\mathbb P}}}
\nc{\QQ}{{{\mathbb Q}}}
\nc{\UU}{{{\mathbb U}}}
\nc{\EE}{{{\mathbb E}}}
\nc{\Icoh}{{I^{\rm coh}}}
\nc{\Qca}{{Q_{\rm ss}}}
\nc{\Qcaa}{{Q^{(1)}_{\rm ss}}}
\nc{\Dcaa}{{D^{(1)}_{{\rm ss}\rightarrow}}}
\nc{\Dca}{{D_{{\rm ss}\rightarrow}}}

\nc{\be}{\begin{equation}}
\nc{\ee}{{\end{equation}}}
\nc{\bea}{\begin{eqnarray}}
\nc{\eea}{\end{eqnarray}}
\nc{\<}{\langle}
\rnc{\>}{\rangle}

\nc{\Hom}[2]{\mbox{Hom}(\CC^{#1},\CC^{#2})}
\nc{\rU}{\mbox{U}}

\begin{document}

\title{Quantum Communications Made Easy\texttrademark
\\ Deterministic Models of Bosonic Channels
}

\author{Graeme Smith and John A. Smolin}

\date{\today}

\address{IBM T.J. Watson Research Center, Yorktown Heights, NY 10598, USA}

\maketitle

\textbf{Information theory establishes the ultimate limits on
  performance for noisy communication systems \cite{Shannon48}.  An
  accurate model of a physical communication device must include
  quantum effects, but typically including these makes the theory
  intractable. As a result communication capacities are not known,
  even for transmission between two users connected by an
  electromagnetic waveguide subject to gaussian noise.  Here we
  present an exactly solvable model of communications with a fully
  quantum electromagnetic field.  This allows us to find explicit
  expressions for all the point-to-point capacities of a noisy quantum
  channel, with implications for quantum key distribution, and fiber
  optical communications.  We also develop a theory of quantum
  communication networks by solving some rudimentary quantum networks
  for broadcasting and multiple access.  When possible, we compare the
  predictions of our new model with those of the orthodox quantum
  gaussian model and in all cases we find capacities in agreement to
  within a constant number of bits.  Thus, in the limit of high signal
  to noise ratios our simple model captures the relevant physics of
  gaussian models while remaining amenable to
  detailed analysis.}\\

\section{Introduction}
A fundamental property of any communication system is the maximum rate
of data transmission possible using the best communication schemes.
This is called the {\em capacity} of a channel.  It is usually
calculated as a function of noise levels and subject to a limited
power budget.  In 1948, Shannon \cite{Shannon48} presented a beautiful
theory of information both formulating and solving the capacity
problem.  For the specific case of a channel with additive white
gaussian noise his formula can be solved explicitly giving the
classical capacity $C$ as a function of the signal to noise ratio
(SNR) as
\begin{equation}
C({\rm SNR})=\frac{1}{2} W\log (1+{\rm SNR}) .
\end{equation}
This formula guided the development of practical schemes that are now
in use, culminating in efficient codes that
come close to achieving the theoretical limit \cite{RU03}.

Noise is not a purely mathematical abstraction, but must arise from
some physical process.  Such processes are properly described, of
course, by quantum mechanics, and therefore calculating the true
information carrying capacity of a channel requires a
quantum-mechanical treatment.  It is also natural, then, to consider
new types of capacities, such as the capacity of a channel for
transmitting quantum states coherently (the \emph{quantum capacity}),
or classical states securely (the \emph{private capacity}).
Unfortunately, unlike in classical information theory, for most
quantum channels none of these capacities are known (see, for for
example, \cite{SS96,DSS98} for the quantum capacity and
\cite{Hastings} for the classical capacity, and \cite{RSS08} for the
private capacity).

Whereas much of the existing work on quantum channels has concentrated on
abstract finite-dimensional channels, here we would like to study the
problem in a more realistic setting.  Our method is well suited to
channels consisting of gaussian noise in bosonic electromagnetic
modes, though it is substantially more general.  We will be able to
calculate classical, quantum, and private capacities for a wide range
of realistic channels.

Because quantum information cannot be cloned \cite{WZ82}, knowledge
gained by the environment about a signal is necessarily detrimental to
quantum transmission.  This need to consider what information is
transmitted to the environment as well as what goes to the intended
receiver puts an analysis of quantum capacity on par with the study of
the classical multi-user broadcast channel (which is also notoriously
difficult to analyze \cite{Cover98}).  To make the problem tractable
we, following the work of \cite{ADT07}, substitute a discretized and
deterministic model for the actual channel.  In the limit of high SNR
this model will capture the important features of the real channel,
and allow us to calculate capacities to within a small number of bits.
The general approach can be thought of as discretizing the continuous
system under consideration in a very simple way and then truncating
signals smaller than the noise power.  The result is a deterministic
channel model that is easy to analyze.

\section{Additive Gaussian Channels}

\begin{figure}[htb]
\centering
\subfloat[]{\includegraphics[height=3in]{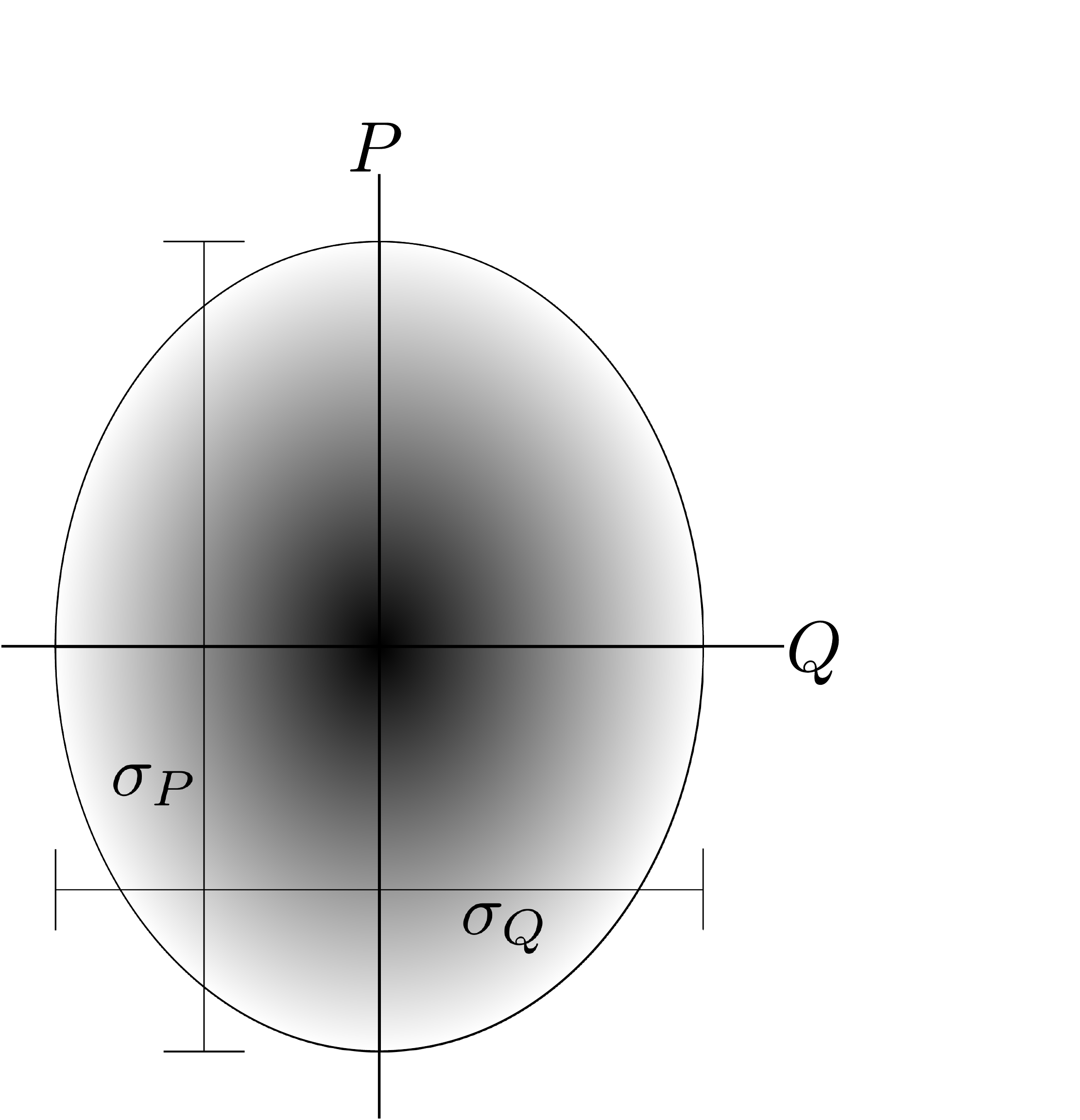}} 
\subfloat[]{\includegraphics[height=3in]{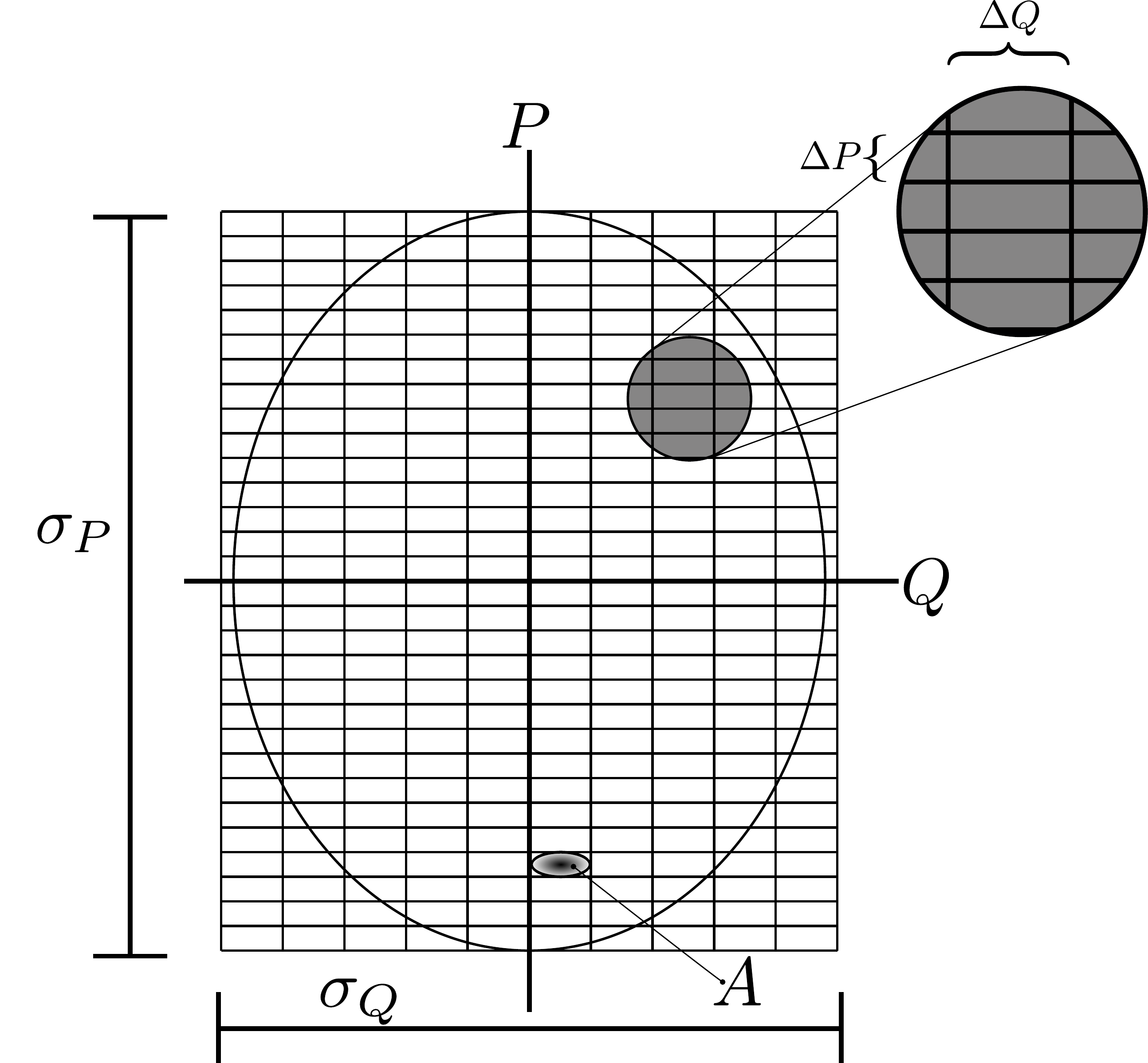}}

\caption{Phase-space representation of states (Wigner Functions): (a)
  Wigner function of a gaussian mixed state with variances $\sigma_Q$
  and $\sigma_P$.  (b) The discrete quadrature model.  The mixed state
  has been approximately decomposed into nonoverlapping rectangles
  with width $\Delta Q$ and height $\Delta P$.  These rectangles are
  the discrete states of the model which replaces original physical
  system.  We imagine them as approximations to the pure squeezed
  state shown at point $A$. }

\label{fig:Wigner}
\end{figure}

States of electromagnetic modes are described by their quadratures
$P,Q$ and are called {\em gaussian} when they can be completely
characterized by a matrix $\gamma$ containing the covariances of these
quadratures \cite{EW05}.  Such states can be visualized by their {\em
  Wigner functions}---quasi-probability distributions depicting the
state's location in phase space.  Gaussian states have Wigner
functions which are ellipsoids with gaussian profiles.
Fig. \ref{fig:Wigner}(a) shows the Wigner function for a gaussian
state with covariances $\sigma_P,\sigma_Q$.  The minimum uncertainty
state has $\sigma_P \sigma_Q = 1/2$ (in units of $\hbar$).  Such
states are always pure.  Mixed states have $\sigma_P \sigma_Q > 1/2$
and can be thought of as mixtures of pure states.

We will replace the quadratures $P,Q$ with a simpler discretized model
described in Fig. \ref{fig:Wigner}(b).  We call this the
\emph{discrete quadrature} (DQ) model.  Roughly speaking, states in
phase space are more distinguishable when their Wigner functions are
less overlapping.  To reflect this property, we divide phase space up
into a lattice of nonoverlapping rectangles that we take to be
perfectly distinguishable.  The smallest physical state has area $1/2$.
The Wigner function is thus replaced with several rectangles of 
area $1/2$ which tile the region of phase space where it is nonnegligable.
The associated state in our model is a uniform mixture of these
distinguishable rectangles.

The model of the action of a channel model will also be discrete: Since
all the output states are perfectly distinguishable rectangles, every
pair of input states will either be mapped to distinguishable outputs,
or to the same output.  Calculating the achieved communication rate is
then a simple matter of counting the distinguishable output states.  A
full calculation of the capacity will also include a maximization over
modulation schemes, \emph{i.e.}  the set of input states employed.
Note that not all modulation schemes are physical.  The choice must
obey the following constraints:
\begin{itemize}
\item $\frac{1}{2}(\sigma_Q^2 + \sigma_P^2) \leq W$ (finite power)
\item $\Delta P \Delta Q \ge 1/2$  (quantum uncertainty)
\item $\Delta P \le \sigma_P,$ $\Delta Q \le \sigma_Q$ (common sense)
\end{itemize}

\begin{figure}[htbf]
\includegraphics[scale=0.4]{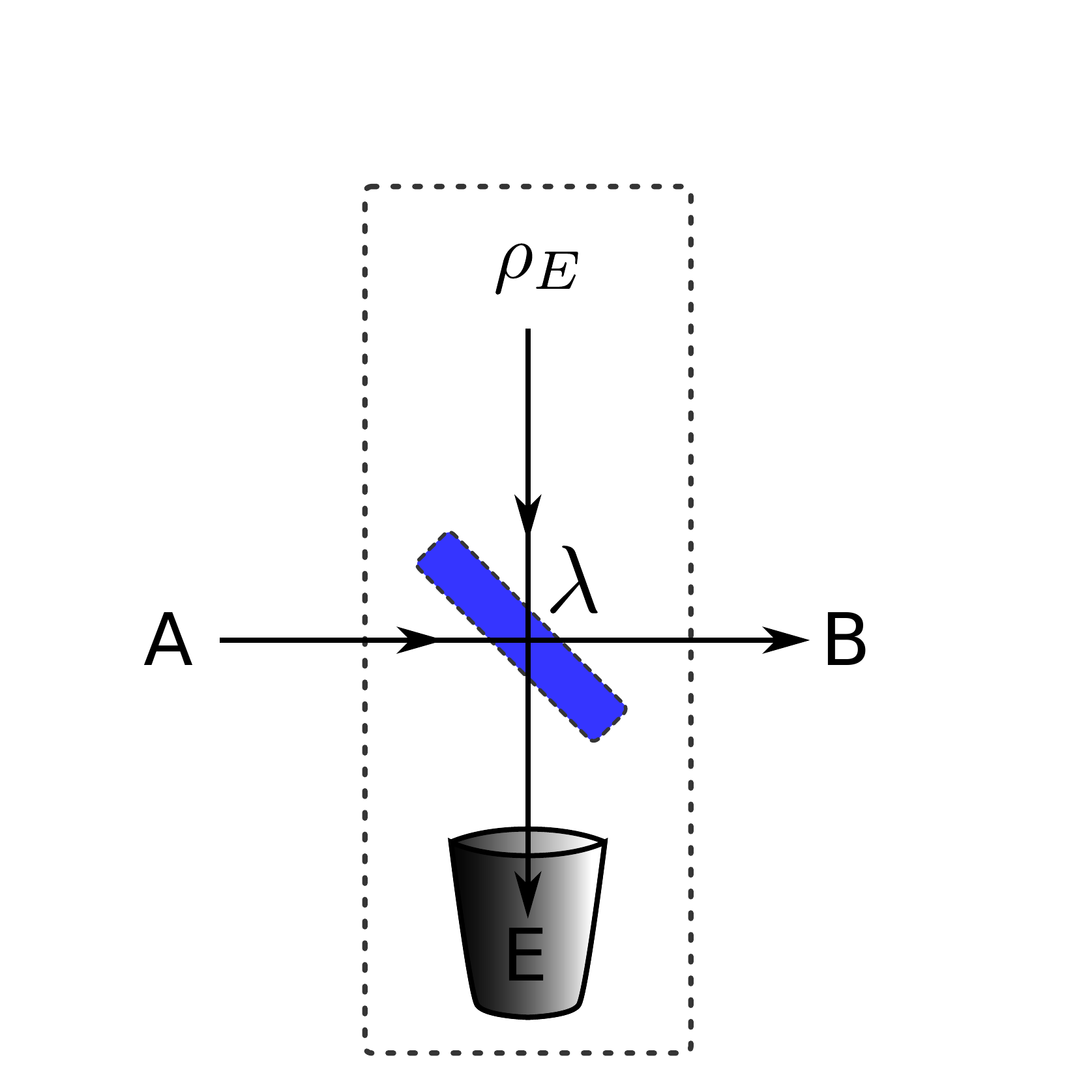}
\caption{Additive gaussian noise noise channel.  An input state in beam A
is combined with an environmental input $\rho_E$ on a beamsplitter
with transmitivity $\lambda$.  One output beam is discarded (remains available
only to the environment) and the other is the channel output B.
When the environment's
input $\rho_E$ is a vacuum state, this gives the attenuation channel.  When
$\rho_E$ is a thermal state, it is the thermal noise channel. 
\label{thermalchannel}}
\end{figure}

The evolution of bosonic states under the action of channels is
described by the evolution of the system's quadratures.  For gaussian
noise, which arises from quadratic Hamiltonian interaction with
gaussian environment modes, the allowed evolutions take a particularly
simple form, being completely described by a symplectic matrix
\cite{HW01}.  For example, $P\rightarrow \sqrt{\lambda} P +
\sqrt{1-\lambda}\ r,\ Q\rightarrow \sqrt{\lambda} Q +
\sqrt{1-\lambda}\ s$, results from interaction with an environment
mode described by $r,s$ (see Fig. \ref{thermalchannel}).  The \emph{additive
gaussian noise channel} arises when the environment begins in a
gaussian state, and the \emph{thermal noise channel},
when it is initially in a thermal state.

\subsection{Classical Capacity}

\begin{figure}[htbf]
\includegraphics[scale=0.2]{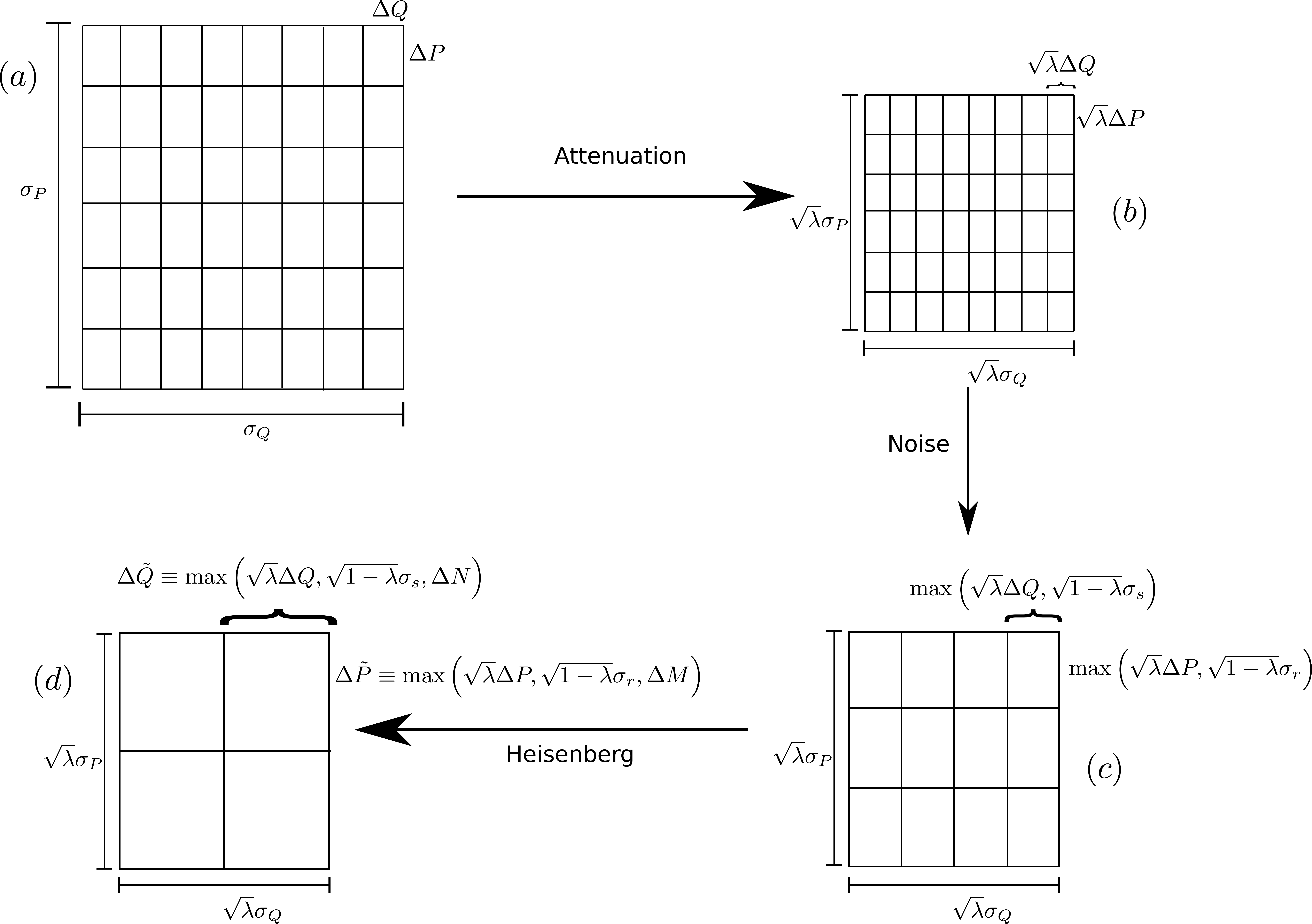}
\caption{The DQ model of the action of a channel.  (a) The input space
is divided up into tiles of $\Delta Q \times \Delta P$.  (b) Attenuation
has shrunk the tiles to $\sqrt{\lambda} \Delta Q \times \sqrt{\lambda} \Delta P$.  (c)  Noise has redenered some tiles confusable, so the effective tile size is increased.  (d) If any tiles are smaller than the smallest tile size
allowed by uncertainty, the tiles are enlarged again.
\label{actionfigure}
}
\end{figure}

\subsubsection{General Additive Gaussian Noise}
\begin{figure}
\centering
\subfloat[]{\includegraphics[height=1.5in]{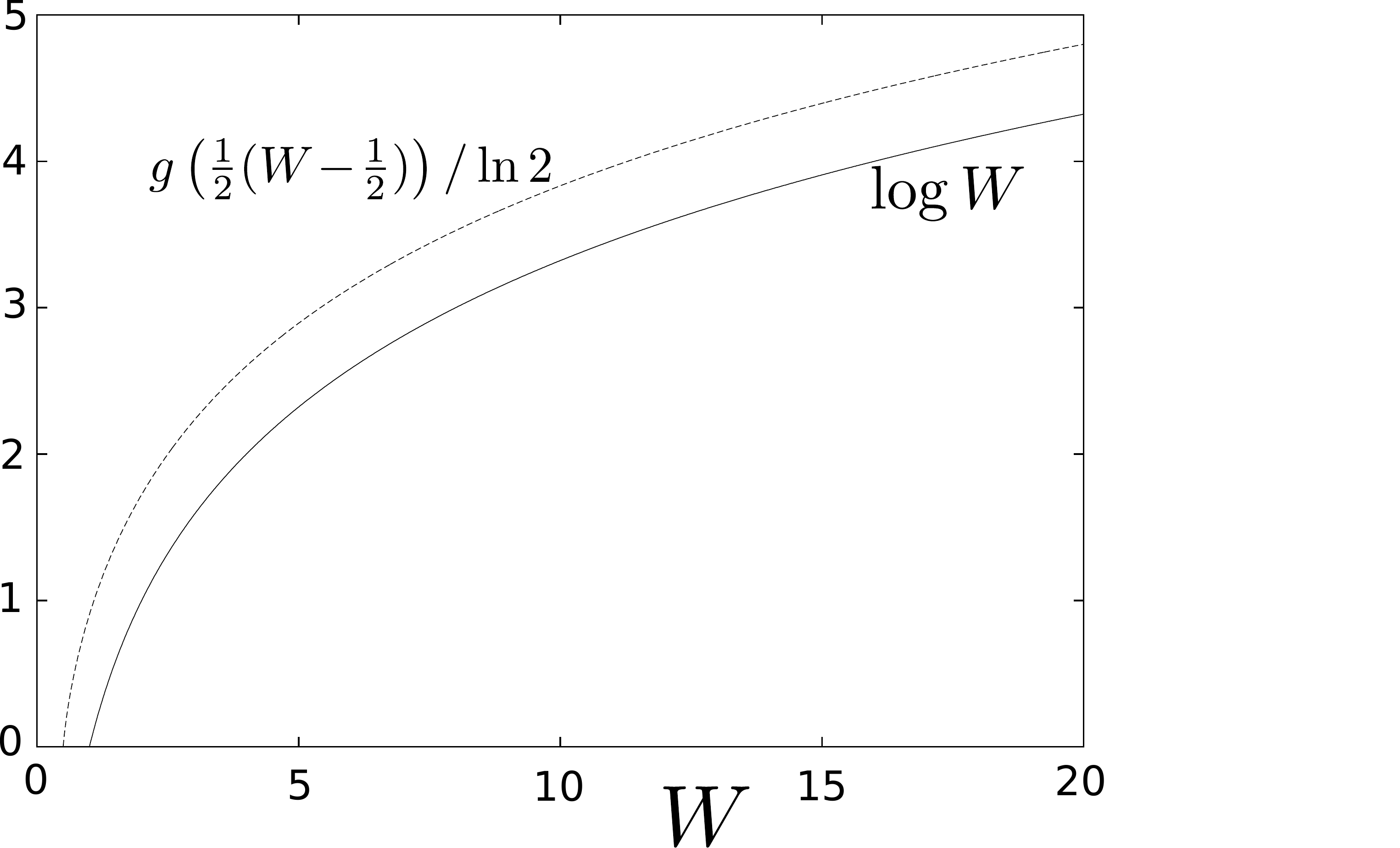}}
\subfloat[]{\includegraphics[height=1.5in]{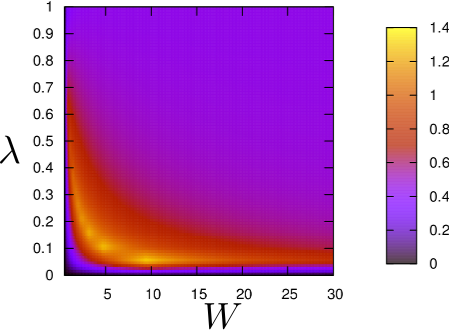}}
\caption{The classical capacity of the attenuation channel.  (a) For $\lambda=1/2$, the
  actual capacity $g((W-1/2)/2)/\ln 2$ and the capacity calculated using the discrete
  quadrature model $\log W$.  (b) The difference between the
  actual and discrete quadrature capacity as a function of both
  $\lambda$ and power.  Note that the difference is never greater
  than 1.4 bits.
\label{classicalcapacityattenuation}
}
\end{figure}

We now apply our discretization procedure to model classical
communication over an additive gaussian noise channel.
Fig. \ref{actionfigure} illustrates the following analysis.  We
suppose that the input signal has some average power constraint $W\geq
\frac{1}{2}(\sigma_P^2+\sigma_Q^2)$.  Furthermore, we fix a modulation
scheme, to be optimized over later, which amounts to deciding the
shape of the rectangles in the discretization shown in
Fig. \ref{fig:Wigner}b.  Given the variables $\sigma_P,\sigma_Q$ and
rectangle shape $\Delta P,\Delta Q$ with $\Delta P \Delta Q = 1/2$,
our input space has $\sigma_P \sigma_Q / (\Delta P \Delta Q)$
distinguishable states (Fig. \ref{actionfigure}a).  We must now
determine how many distinguishable outputs these get mapped to.
Attenuation by $\lambda$ maps the entire input space to a
$\sqrt{\lambda}\sigma_P \times \sqrt{\lambda} \sigma_Q$ rectangle of
$\sqrt{\lambda} \Delta P \times \sqrt{\lambda} \Delta Q$ tiles
(Fig. \ref{actionfigure}b).  When a noise of typical size
$\sqrt{1-\lambda} \sigma_r$ is added to the $P$ quadrature, tiles
closer than this are taken to be confusable, and similarly for $Q$,
which leads us to ``meta''-tiles of dimension
\begin{equation}
\max(\sqrt{\lambda}\Delta P,\sqrt{1-\lambda} \sigma_r)\times \max(\sqrt{\lambda} \Delta Q,\sqrt{1-\lambda} \sigma_s)\ {\rm (Fig.\ \ref{actionfigure}c).}
\end{equation}
If it happens that this rectangle's area is less than $1/2$
the tiles in Fig. \ref{actionfigure}c are smaller than the minimum
tile size allowed by uncertainty.  We must then chose a tile shape,
$\Delta N \times \Delta M$ with $\Delta N \Delta M = 1/2$ satisfying
uncertainty.  We thus find a final tile dimension of
$\Delta\tilde{P}\times \Delta \tilde{Q}$ with
\begin{align}
\Delta\tilde{P}&\equiv \max(\sqrt{\lambda}\Delta P,\sqrt{1-\lambda} \sigma_r,\Delta N)\\
 \Delta\tilde{Q}&\equiv \max(\sqrt{\lambda} \Delta Q,\sqrt{1-\lambda} \sigma_s,\Delta M)\ \ {\rm (Fig.\ \ref{actionfigure}d)}. 
\end{align}
This gives a total number of distinguishable output states of 
$\lambda \sigma_P \sigma_Q/(\Delta \tilde{P}\Delta \tilde{Q})$
and a classical capacity $\mathbf{C}$ of 
\begin{equation}
\mathbf{C}=\max\log\left(\frac{\lambda \sigma_P \sigma_Q}
{\Delta \tilde{P}\Delta \tilde{Q}}\right)\ 
\label{capacity}
\end{equation}
where the maximization is over all of the constrained variables:
$\frac{1}{2}(\sigma_P^2+\sigma_Q^2)\leq W,\  \Delta P\Delta Q \geq \frac{1}{2},\ 
\Delta P \leq \sigma_P,\ \Delta Q \leq \sigma_Q,\ \Delta N\Delta M \geq \frac{1}{2}$.
We now evaluate the preceding formula for some important special cases.

\subsubsection{Example: Attenuation}
The attenuation channel is an additive channel with transmissivity
$\lambda$ and pure environment in a vacuum state with $\sigma_r
=\sigma_t =\frac{1}{\sqrt {2}} $. To good approximation, this channel
describes the propagation of the signal through lossy optical fiber.
Letting $\Delta P =\Delta Q =\frac{1}{\sqrt{2}} $ and $\Delta M
=\Delta N =\frac{1}{\sqrt{2}} $, minimizes $\Delta
\tilde{P}\Delta\tilde{Q}=\frac{1}{2} $ while satisfying the
constraints. This leads to a capacity of $\log\lceil 2W\lambda\rceil
$. For this channel, we actually know the classical capacity exactly
\cite{Giovannetti}.  Indeed, the true capacity $g\left(\lambda (W
-\frac{1} {2})\right)/\ln 2$, where $g(x)=(x+1)\ln (x+1)-x\ln x$, differs from
our estimate by no more than 1.4 bits (see Fig. \ref{classicalcapacityattenuation}).

\subsubsection{Example: Classical Noise}

When a channel applies gaussian-distributed kicks in phase space, it
is called a classical noise channel.  This arises as a limiting case
of the thermal noise channel shown in Fig. \ref{thermalchannel} with
$\lambda\rightarrow 1$ and $\sigma_r=\sigma_s\rightarrow \infty$ while
keeping $\sigma_r \sqrt{1-\lambda} = \mu$.  We would like to evaluate
Eq. (\ref{capacity}) for this channel, where we have
$\Delta\tilde{P}=\max(\Delta P,\mu)$ and $\Delta\tilde{Q}=\max(\Delta
Q,\mu)$.

First we consider the case $\mu^2\geq 1/2$.  First note that
$\Delta\tilde{P}\Delta\tilde{Q}\geq \mu^2$.  Choosing $\Delta P =
\Delta Q = 1/\sqrt{2}$ achieves this lower bound.  Note also that
$\max_{W\geq \frac{1}{2}(\sigma_P^2+\sigma_Q^2)} \sigma_P \sigma_Q =
W$ is achieved for $\sigma_P=\sigma_Q=\sqrt{W}$.  Thus by choosing
$\Delta P=\Delta Q=1/\sqrt{2}$ and $\sigma_P=\sigma_Q=\sqrt{W}$ we
simultaneously maximize $\sigma_P\sigma_Q$ and minimize
$\Delta\tilde{P}\Delta\tilde{Q}$ while satisfying all the constraints
in Eq. (\ref{capacity}).  We thus find a capacity of $\log (W/\mu^2)$.
In a similar way, when $\mu^2 < 1/2$, $\Delta\tilde{P}\Delta\tilde{Q}$ is
minimized by $\Delta P=\Delta Q=1/\sqrt{2}$ so that by the same argument
the capacity is now $\log(2W)$.  A lower bound for the capacity of the
classical noise channel with noise power $\mu^2$ is 
$\left(g(W-1/2+\mu^2)-g(\mu^2)\right)/\ln 2$.  This is achieved by
classical displacements of vaccuum states.  This bound is within 1.45 bits of
the true capacity \cite{Koenig} and agrees with the DQ model to
within 1 bit for all $\mu^2$.

\subsubsection{Example: Dephasing}
Another interesting special case is the classical dephasing channel,
which adds classical noise of power $\mu^2$ to the $Q$ quadrature
while leaving $P$ untouched.  Starting from the quantum channel 
$P\rightarrow \sqrt{\lambda} P +
\sqrt{1-\lambda} r$ and $Q\rightarrow \sqrt{\lambda} Q +
\sqrt{1-\lambda} s$, we take the limit $\lambda\rightarrow 1$ with
$(1-\lambda)\sigma_s^2=\mu^2$, $(1-\lambda) \sigma_r^2\rightarrow 0$.  
So we have the channel $P\rightarrow P$ and $Q\rightarrow Q+ \xi$ where
$\xi$ is a classical guassian random variable with variance $\mu^2$.

\begin{figure}
\centering
\subfloat[]{\includegraphics[height=1.7in]{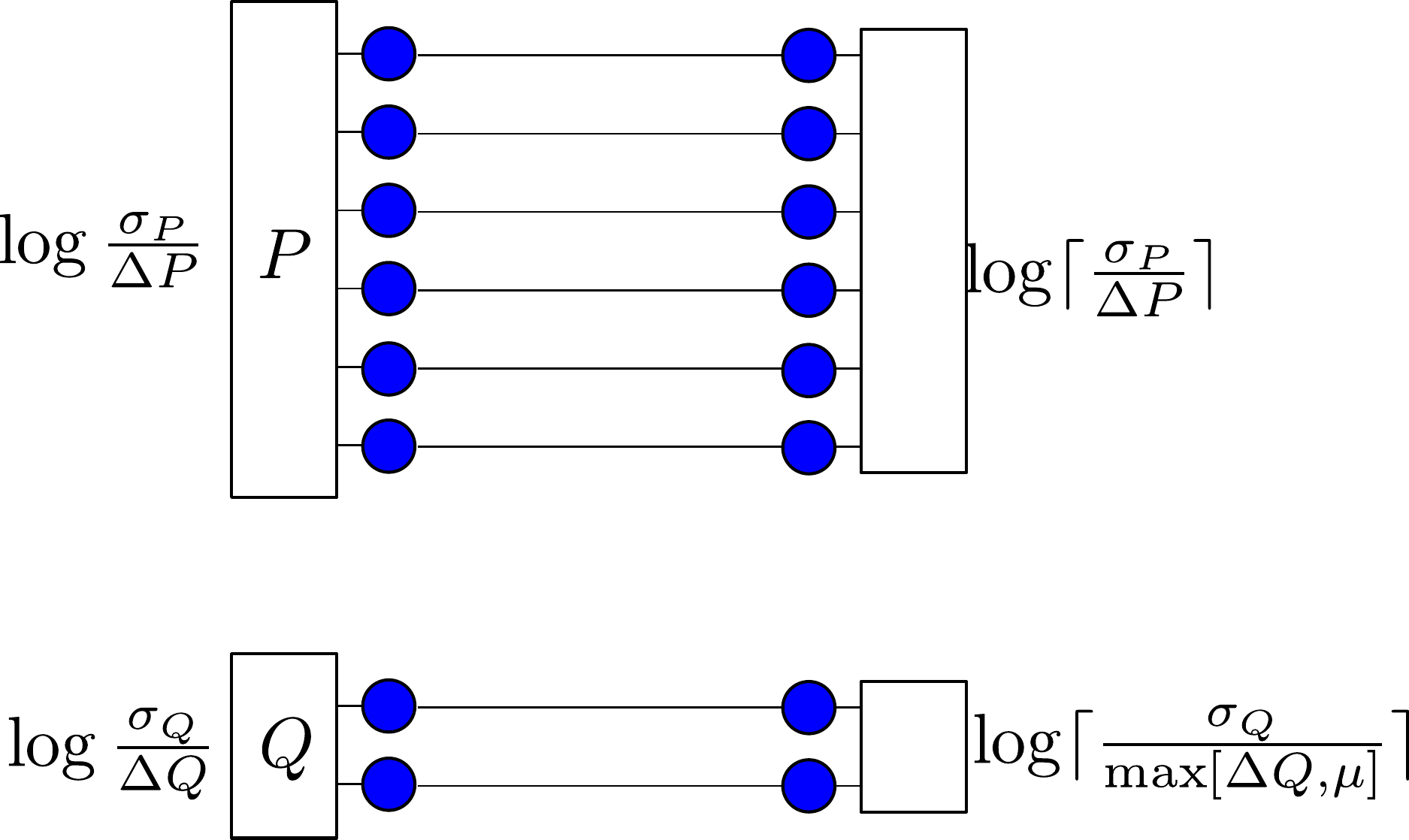}}
\subfloat[]{\includegraphics[height=1.7in]{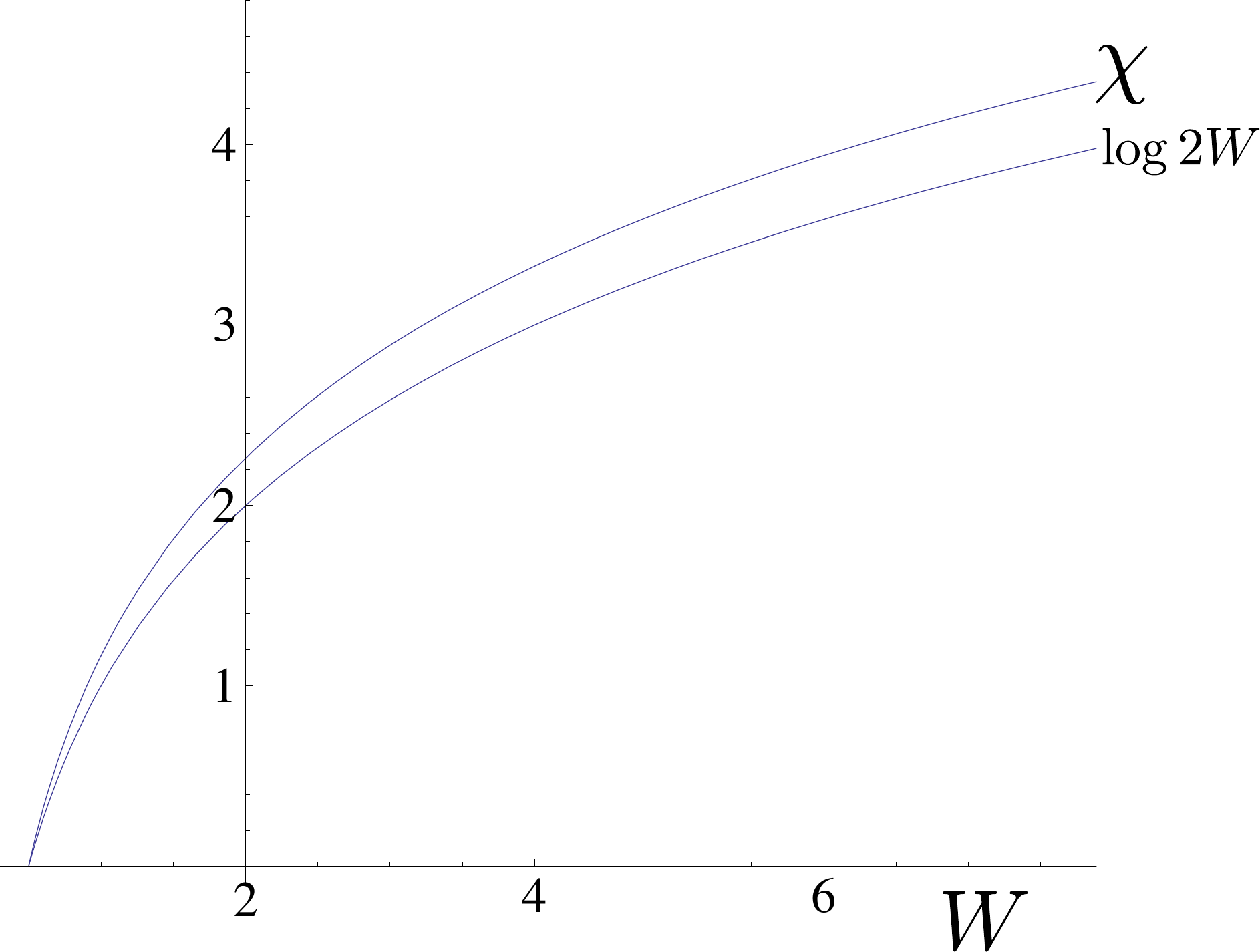}} 
\caption{Dephasing channel.
(a) Discrete quadrature model. The $\max[{\Delta Q,\mu}]$ in the denomiator
of the $Q$ quadrature output takes into account that you cannot distinguish
states smaller than either the size of the input or the size of the applied
classical noise. The `ceiling' brackets inside the logarithms
avoid them blowing up if the noise is so high that there are fewer than
one distinguishable states by this calculation.  (b) Classical capacity of 
the dephasing channel 
with $\mu=1$.  The capacity is achieved with $\Delta P=1/(2 \sqrt{W})$
and $\Delta Q=\sqrt{W}$ and $\sigma_P=\sigma_Q=\sqrt{W}$.  The squeezing and therefore the capacity is limited
only by the input power.
The lower line in the plot is the prediction from the discrete 
quadrature model, the upper is an exact calcuation of the Holevo quantity
given a signal ensemble of maximally-squeezed pure states subject to the power
constraint, which is an achievable rate 
$\chi=\left(g( \frac{2 W \sqrt{1+\mu^2/(2 W)}-1}{2})-g(\frac{\sqrt{1+\mu^2/(2 W)}-1}{2})\right)/\ln 2$.  Our model suggests this is not 
far from the actual capacity.
\label{dephasing}
}
\end{figure}

The discrete quadrature model of this channel is considered in
Fig. \ref{dephasing} , where we have obtained a capacity estimate of
$\log 2W$, independent of noise level.  The modulation scheme to
achieve this involves highly squeezed signal states.  By choosing
states very narrow in $P$ and broad in $Q$, we can pack nearly all of
the signal into the noiseless quadrature.  Thus, for classical
transmission, such a channel achieves rates just as high as in the
noiseless channel, albeit with more difficult modulation.  This
prediction is borne out by computing the Holevo information on the
suggested squeezed ensemble, which an gives achievable rate that
differs from $\log 2W$ by no more than a bit \cite{Hirota,HW01}.

\subsection{Quantum Capacity}
As mentioned above, evaluating the quantum capacity $\mathbf{Q}$
requires assessing not only signals sent from sender to receiver, but
also how information is leaked from sender to the channel's
environment.  While this makes our calculations more complex, we will
nevertheless find simple and reliable estimates for quantum
capacities.  The quantum capacity will be computed as the number of
distinguishable states communcated to the channel's output about which
the environment knows nothing at all.  This complete lack of knowledge
enables one to communicate quantum superpositions of these
distinguishable states, therefore our model imagines
them to define the basis states of a Hilbert space that will be
successfully transmitted.  Note that this is exactly the
definition of the \textit{private capacity} \cite{D03}, and therefore our model
will not be able to separate quantum and private capacities.

\subsubsection{Example: Thermal Noise}

\begin{figure}
\centering
\subfloat[]{\includegraphics[height=1.7in]{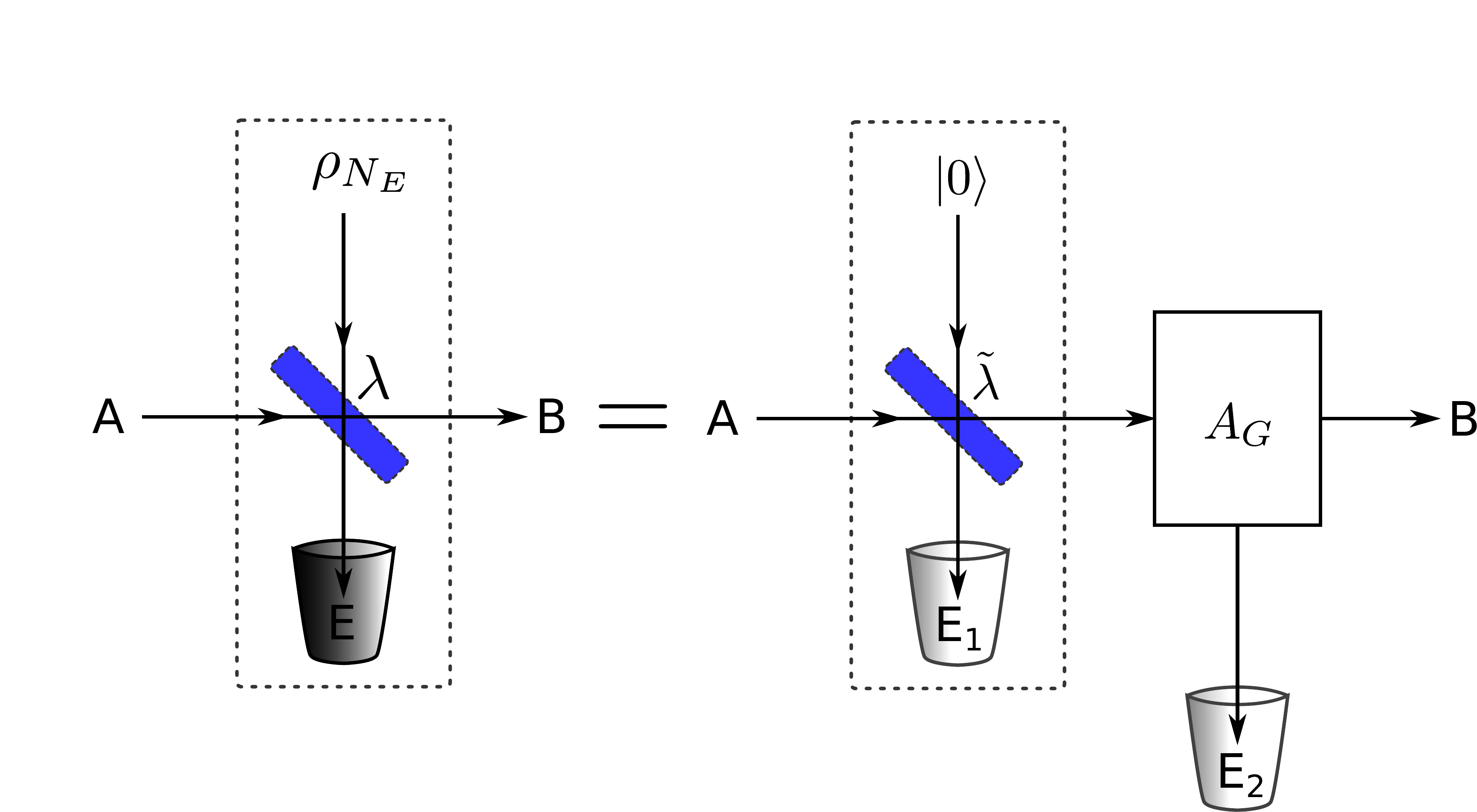}} 
\subfloat[]{\includegraphics[height=1.7in]{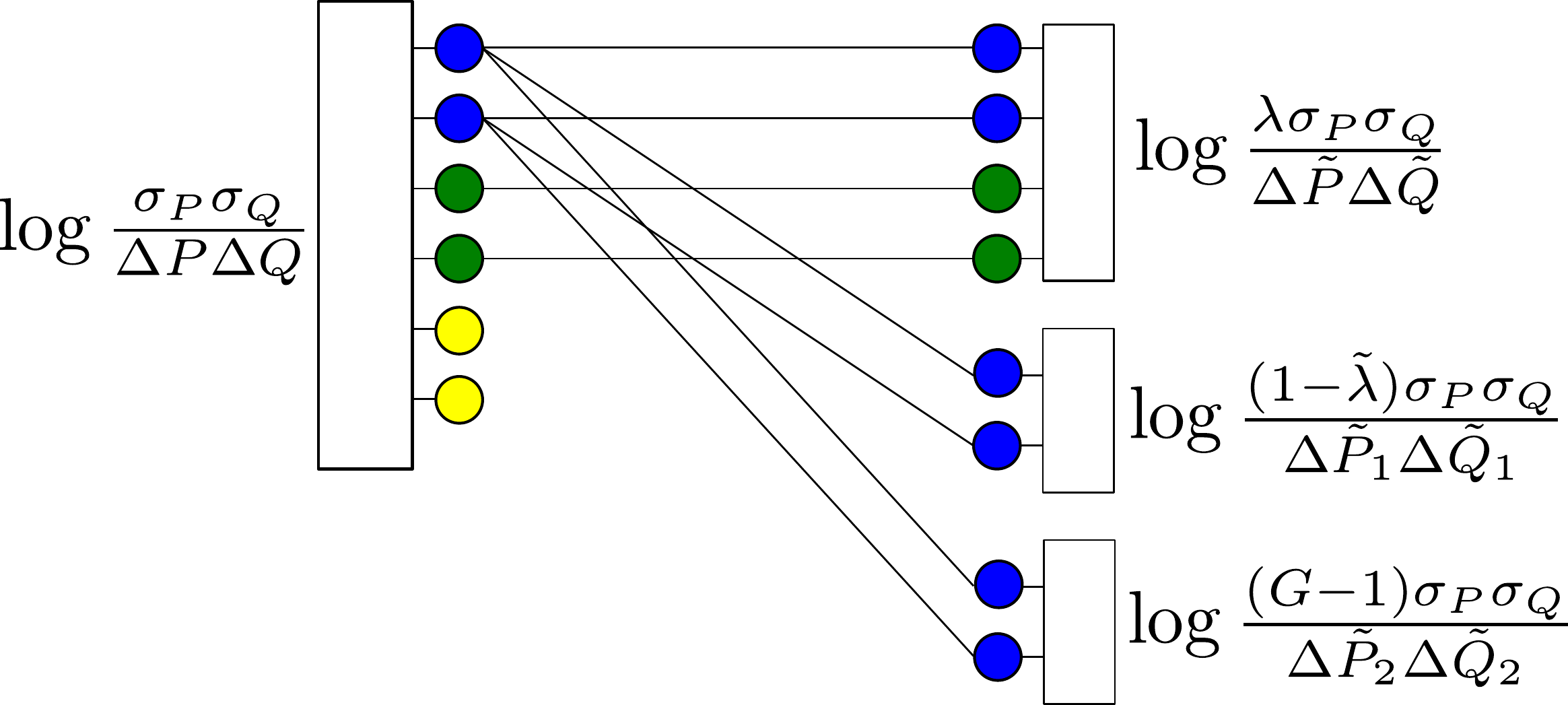}} 
\caption{
(a) Decomposition of thermal noise channel with average photon number $N_E$ into a pure attenuation channel with $\tilde{\lambda}=\lambda/G$ and a
gain channel with gain $G$.  (b) DQ model of this decomposition. 
\label{additivenoisemodel}
}
\end{figure}

We now turn our attention to the thermal noise channel, which maps
$P\rightarrow \sqrt{\lambda} P +\sqrt{1 -\lambda}\ r $ $Q\rightarrow
\sqrt{\lambda} Q+\sqrt{1 -\lambda}\ s $, where $r,s$ are the quadratures of
a thermal state with average photon number $N_E$.  Such a channel can be
decomposed as a composition of a pure loss channel with an ideal amplifier
as shown in Fig. \ref{additivenoisemodel}a.  This decomposition allows us to express the environment's 
state as 
\begin{align}
\label{firstenvironment}\sqrt{1-\tilde{\lambda}} (P,Q)-\sqrt{\tilde{\lambda}} (r_1,s_1)\\
\label{secondenvironment}\sqrt{G}(r_2,s_2)+\sqrt{G-1}(P,-Q)
\end{align}
where $G=(1-\lambda)N_E+1$ and $\tilde{\lambda}=\lambda/G$ and $r_1,s_1$ and
$r_2,s_2$ are the quadratures of two independent vacuum states.

Following our prescription, an input with power $W$ and variances
$\sigma_P^2+\sigma_Q^2 \leq W$ we have $\lambda\sigma_P\sigma_Q/\Delta
\tilde{P}\Delta \tilde{Q}$ distinguishable states at the output, or
\begin{equation}
S_\textrm{output}=\log(\lambda\sigma_P\sigma_Q/\Delta \tilde{P}\Delta \tilde{Q})
\end{equation}
bits.  The highest order bits of the input get mapped to
$\log(\lambda \sigma_P \sigma_Q/\Delta\tilde{P}\Delta\tilde{Q})$
states at the output.  Similarly, there are two environmental modes.
Similarly, according to Eq. (\ref{firstenvironment}), the first
environmental mode gets
\begin{equation}
S_1=\log\left((1-\tilde{\lambda})\sigma_P\sigma_Q/\Delta \tilde{P_1}\Delta \tilde{Q_1}\right)=\log\left(2(1-\lambda/G)\sigma_P\sigma_Q\right)
\end{equation}
bits where 
$\Delta\tilde{P_1}=\max(\sqrt{1-\tilde{\lambda}} \Delta P,\Delta M_1,\sqrt{\tilde{\lambda}} \sigma_{r_1})$ and
$\Delta\tilde{Q_1}=\max(\sqrt{1-\tilde{\lambda}} \Delta Q,\Delta N_1,\sqrt{\tilde{\lambda}} \sigma_{s_1})$.
We always have $\Delta \tilde{P_1} \Delta \tilde{Q_1}\geq 1/2$, and 
$\Delta \tilde{P_1} \Delta \tilde{Q_1}= 1/2$ can be achieved for appropriate $\Delta M_1, \Delta N_1$.

For the second environmental mode, we have $S_2=\log \left((G-1) \sigma_P \sigma_Q/\Delta\tilde{P_2}\Delta\tilde{Q_2}\right)=
\log \left( 2 (G-1) \sigma_P \sigma_Q/G\right)$ where 
$\Delta\tilde{P_2}=\max(\sqrt{G-1} \Delta P,\Delta M_2,\sqrt{G} \sigma_{r_2})$ and
$\Delta\tilde{Q_2}=\max(\sqrt{G-1} \Delta Q,\Delta N_2,\sqrt{G} \sigma_{s_2})$ and 
$\Delta \tilde{P_2} \Delta \tilde{Q_2}= G/2$ is achievable.

Note that the high-order bits transmitted to both mode 1 and mode 2 are identical, but mode 1 gets more of them so that the
total number of bits leaked to the environment is $S_1$.  The quantum capacity is therefore 
\begin{align}
S_\textrm{output}-S_1&=\log(\lambda\sigma_P\sigma_Q/\Delta \tilde{P}\Delta \tilde{Q})- \log\left(2(1-\lambda/G)\sigma_P\sigma_Q\right)\label{eqoutput}\\
&=\log \lambda - \log(1-\lambda/G)   - \log 2 \Delta\tilde{P}\Delta\tilde{Q} \\
\label{capacityQ} &=\log \lambda - \log(1-\frac{\lambda}{(1-\lambda) N_E+1})   - \log 2 \Delta\tilde{P}\Delta\tilde{Q} 
\end{align}

We now evaluate (\ref{capacityQ}) for the attenuation channel
($N_E=0$): As we showed in our discussion of the classical capacity of
the attenuation channel, by appropriate choices of $\Delta P,\Delta Q$
and $\Delta M,\Delta N$, we can achieve
$\Delta\tilde{P}\Delta\tilde{Q}=1/2$.  Then the capacity is
$\log{\lambda}-\log{(1-\lambda)}$ for $\lambda\ge 1/2$ and unlimited
input power.  This is exactly the quantum capacity of the gaussian
attenuation channel \cite{WG2007}.  In fact our model offers slightly
more information.  Throughout in order for our model to make sense,
the estimates of the number of levels transmitted must be integers.
To ensure this in Eq. (\ref{eqoutput}) we need to have power $W
\gtrsim 1/\lambda$, which suggests the minimum power necessary to
achieve capacity.

\section{Simultaneous Quantum and Classical Communication}
\begin{figure}
\centering
\subfloat[]{\includegraphics[height=1.5in]{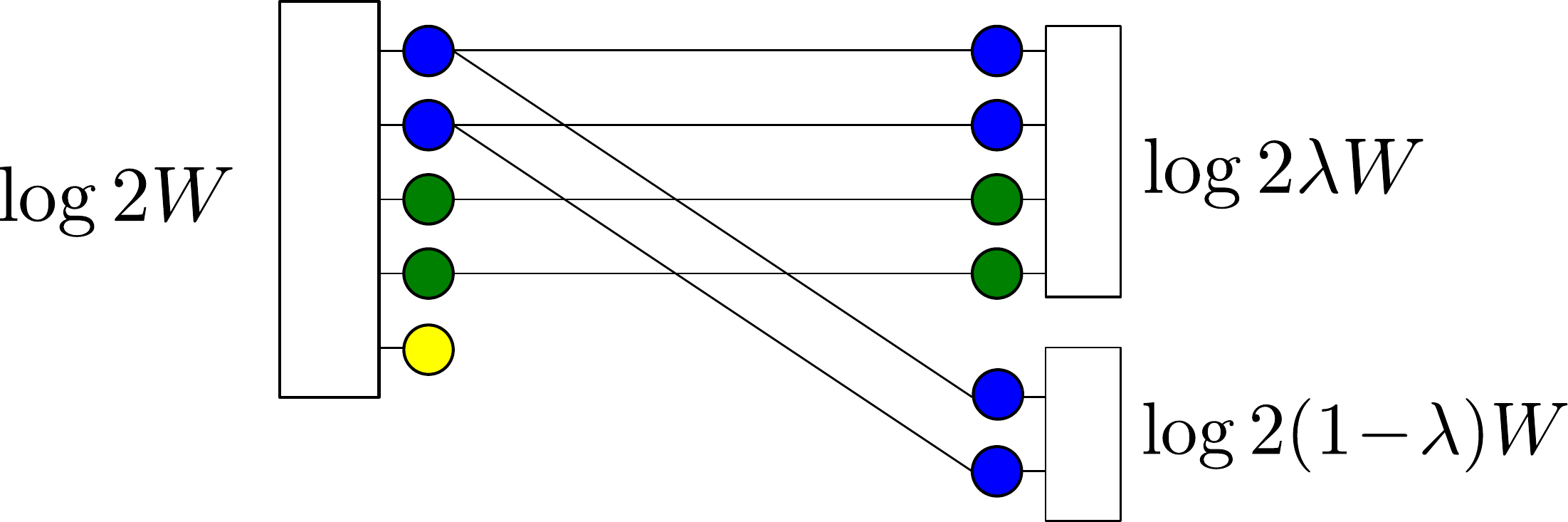}} 
\subfloat[]{\includegraphics[height=1.5in]{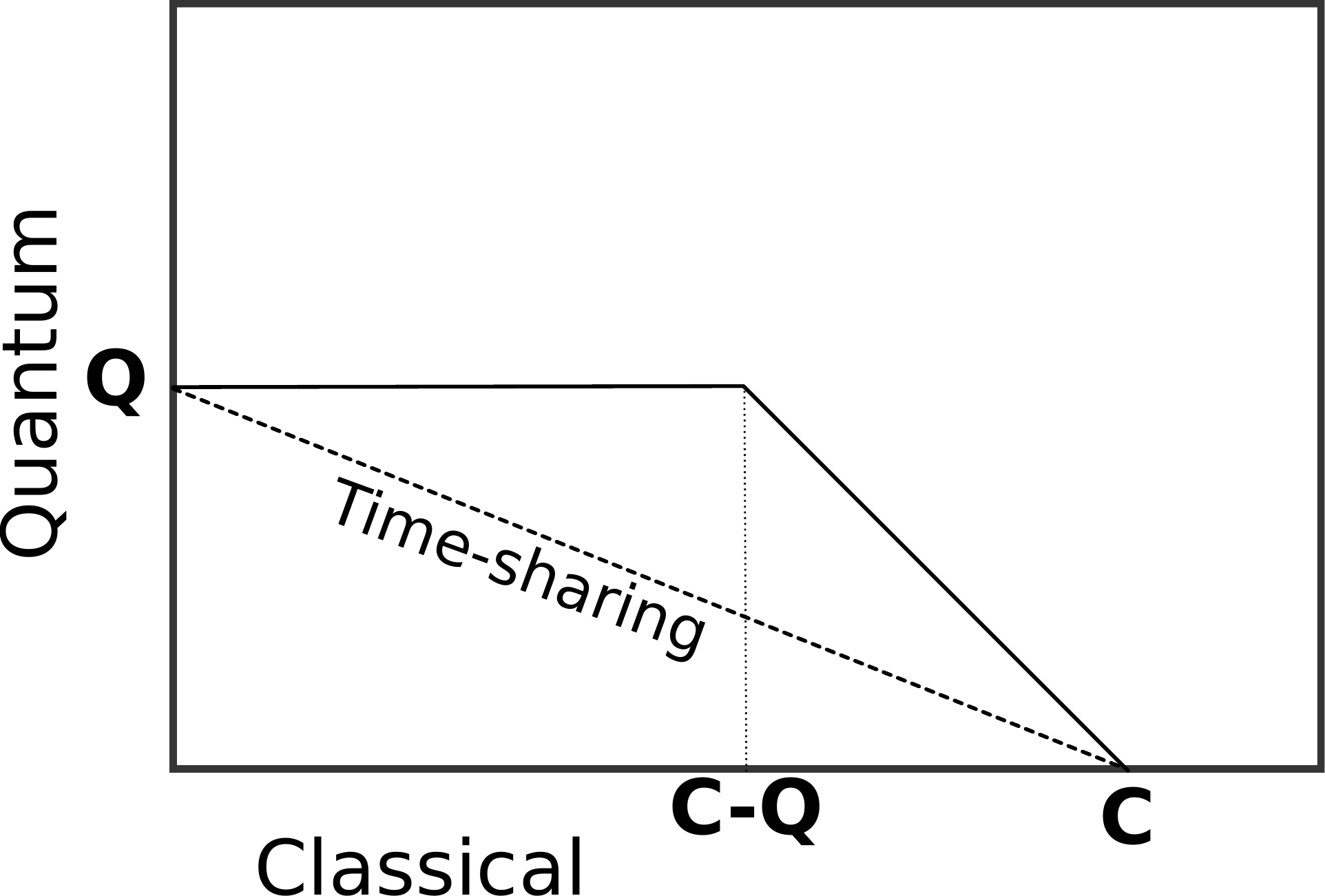}}
\caption{
Simultaneous classical and quantum communication over a gaussian attenuation channel.  (a) DQ Channel model.  The higherst-order (blue) bits are seen by the
environment and therefore can never be used for quantum communication.  They
can always be used for classical communication, even when the lower order bits
are being used for quantum communication.
(b) Rate region compared to the time-sharing tradeoff.
\label{ratebits}
}
\end{figure}

Our model also allows for a simple understanding of the recently
discovered \cite{WHG12} tradeoff between classical and quantum
communication over \textit{the same} optical channel.  Switching
between the two different optimal communication strategies for the two
types of information (\textit{time sharing}), gives a linear tradeoff
between the two communication rates.  In \cite{WHG12}, it was shown
that rate region for an attenuation channel could be substantially
larger than this naive strategy would suggest.  A glance at Figure
\ref{ratebits} makes it clear that while sending quantum information at
capacity, it is possible to simultaneously send some classical
information with no degradation of the quantum transmission.  Only
after sending classical information at a rate greater than $\mathbf{C}-\mathbf{Q}$ does
a linear tradeoff emerge.

\begin{figure}[htbf]
\centering
\subfloat[]{\includegraphics[height=1.5in]{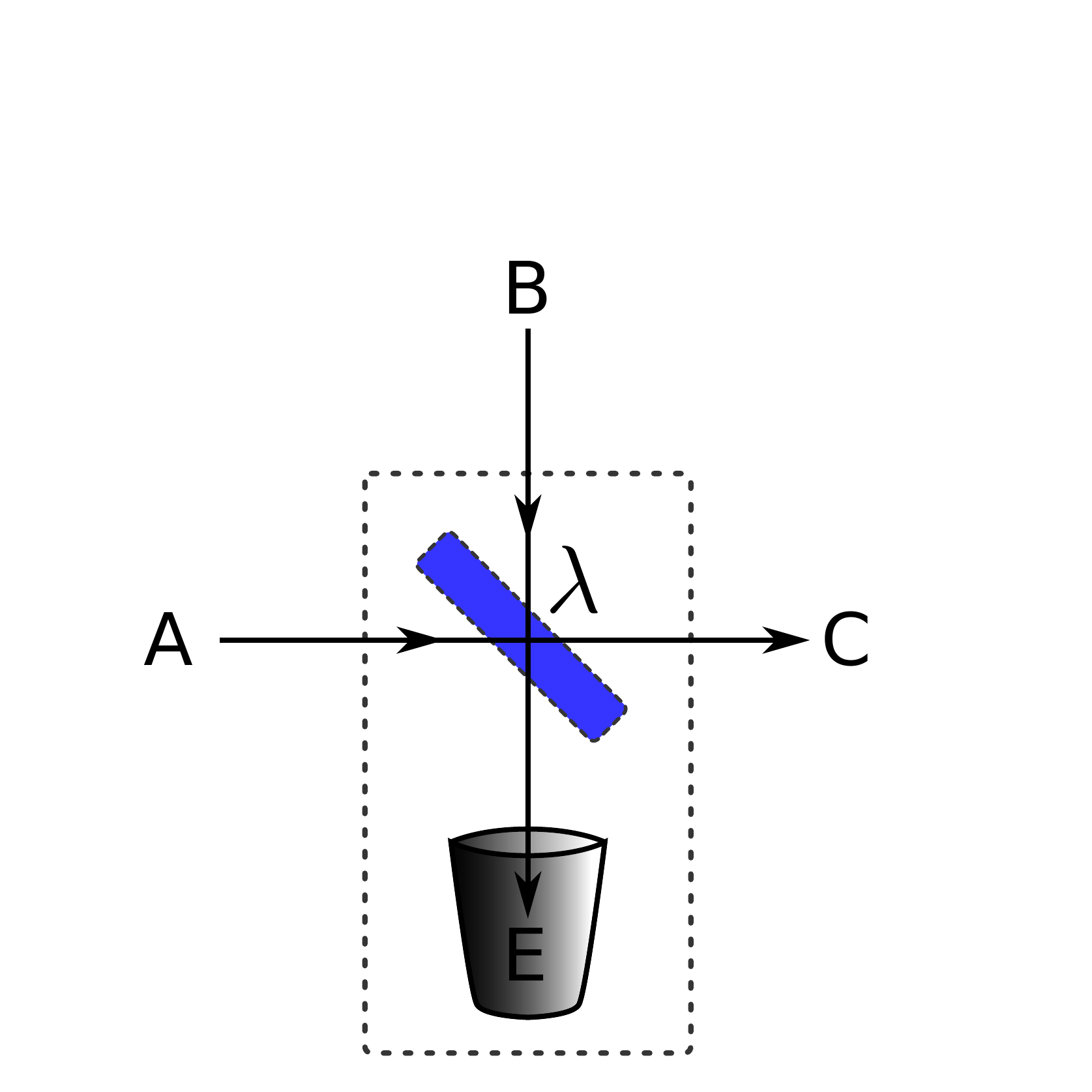}} 
\subfloat[]{\includegraphics[height=1.5in]{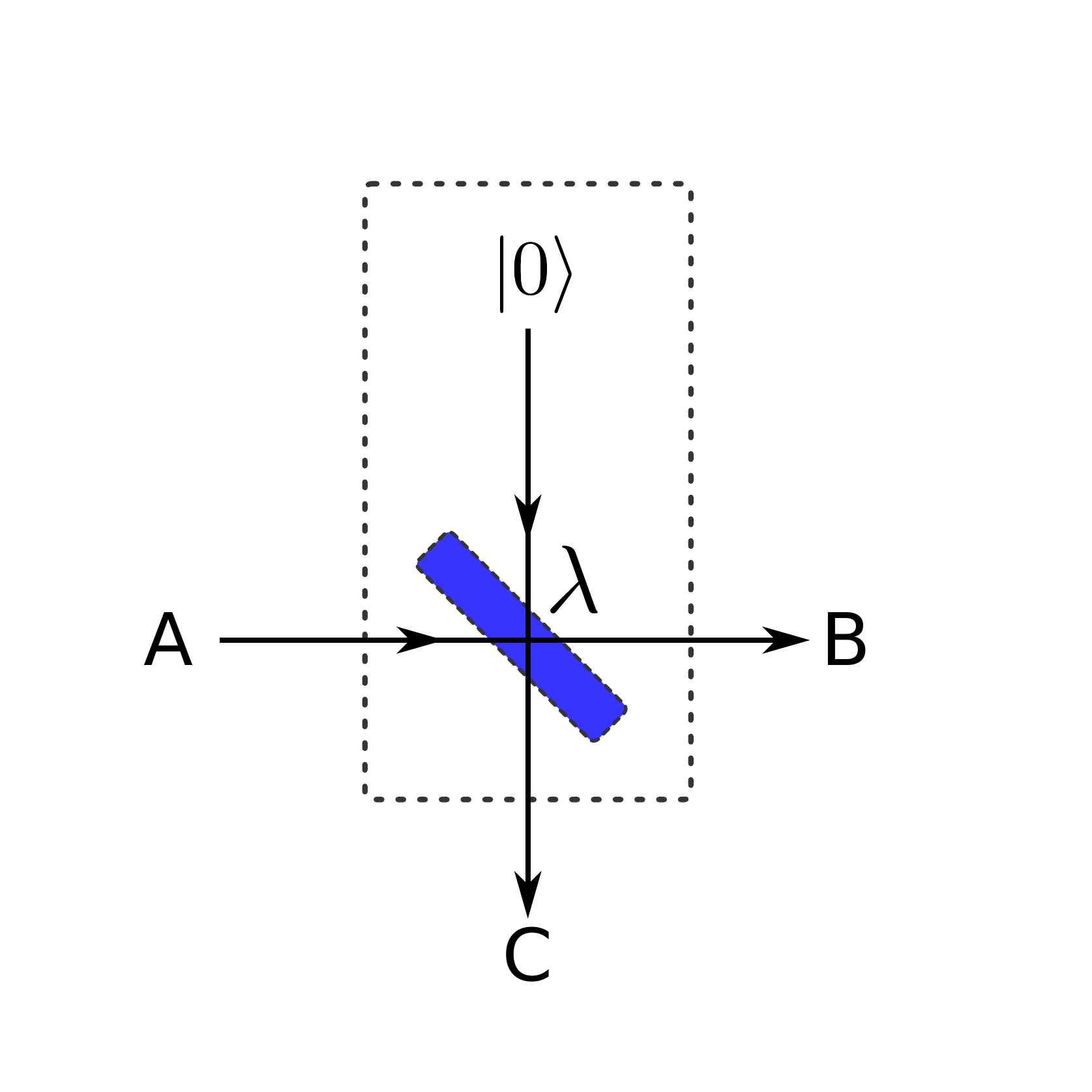}}
\caption{
Multi-user Channels:  (a) Multiple-access channel.  (b) Broadcast channel.  
\label{multiuserchannels}
}
\end{figure}

\section{Multi-User Communication}
A useful and natural generalization of the channel capacity problem is
when there are multiple senders and/or receivers.  Our simple model is
well adapted to extracting useful answers in this setting which is
typically highly intractable, even in the classical setting
\cite{CoverThomas,ElGamalKim}.  Below we consider two multi-user channels: The
multiple access channel with two senders and one receiver and the
broadcast channel with one sender and two receivers (see Fig. \ref{multiuserchannels}).

\begin{figure}[htbf]
\centering
\subfloat[]{\includegraphics[height=1.5in]{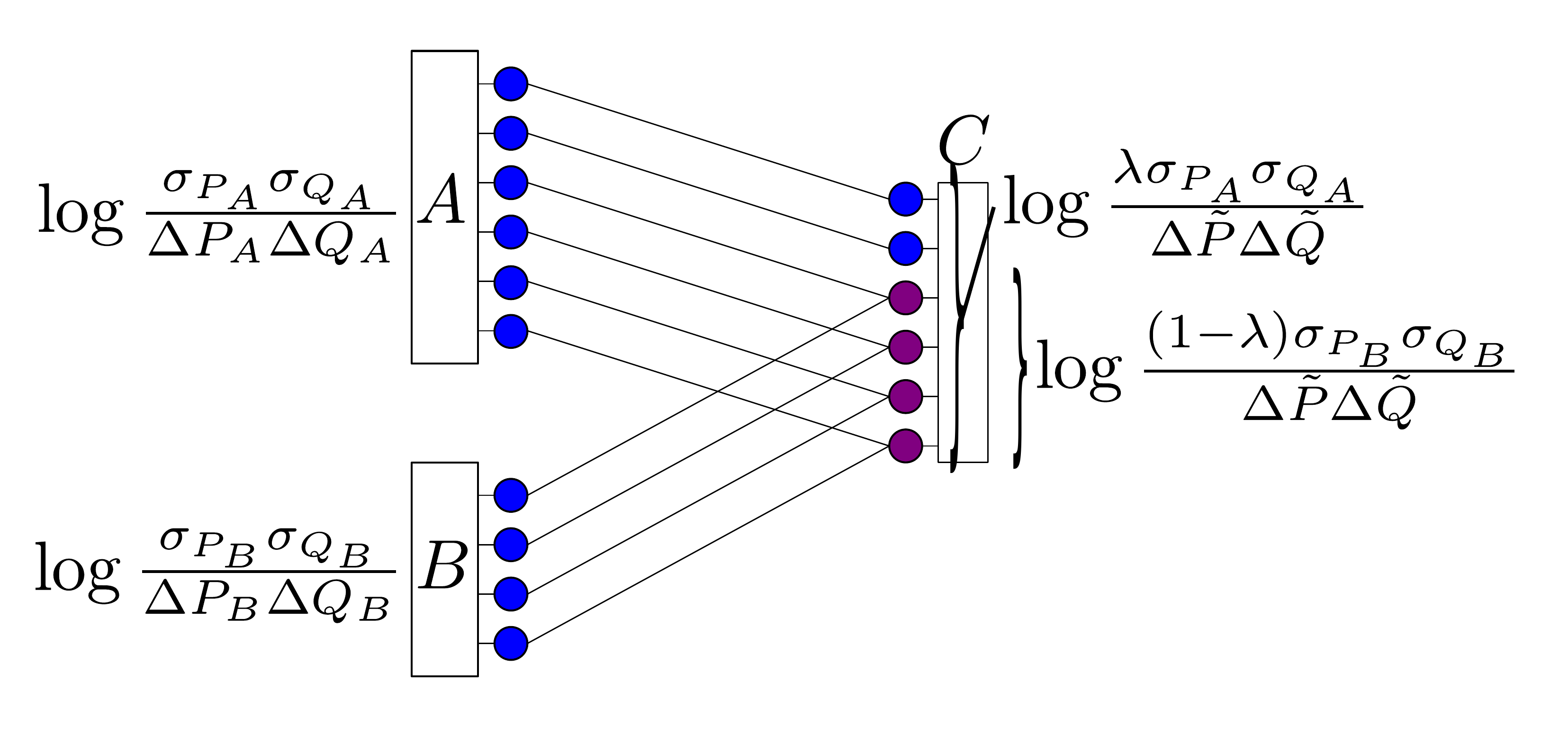}} 
\subfloat[]{\includegraphics[height=1.5in]{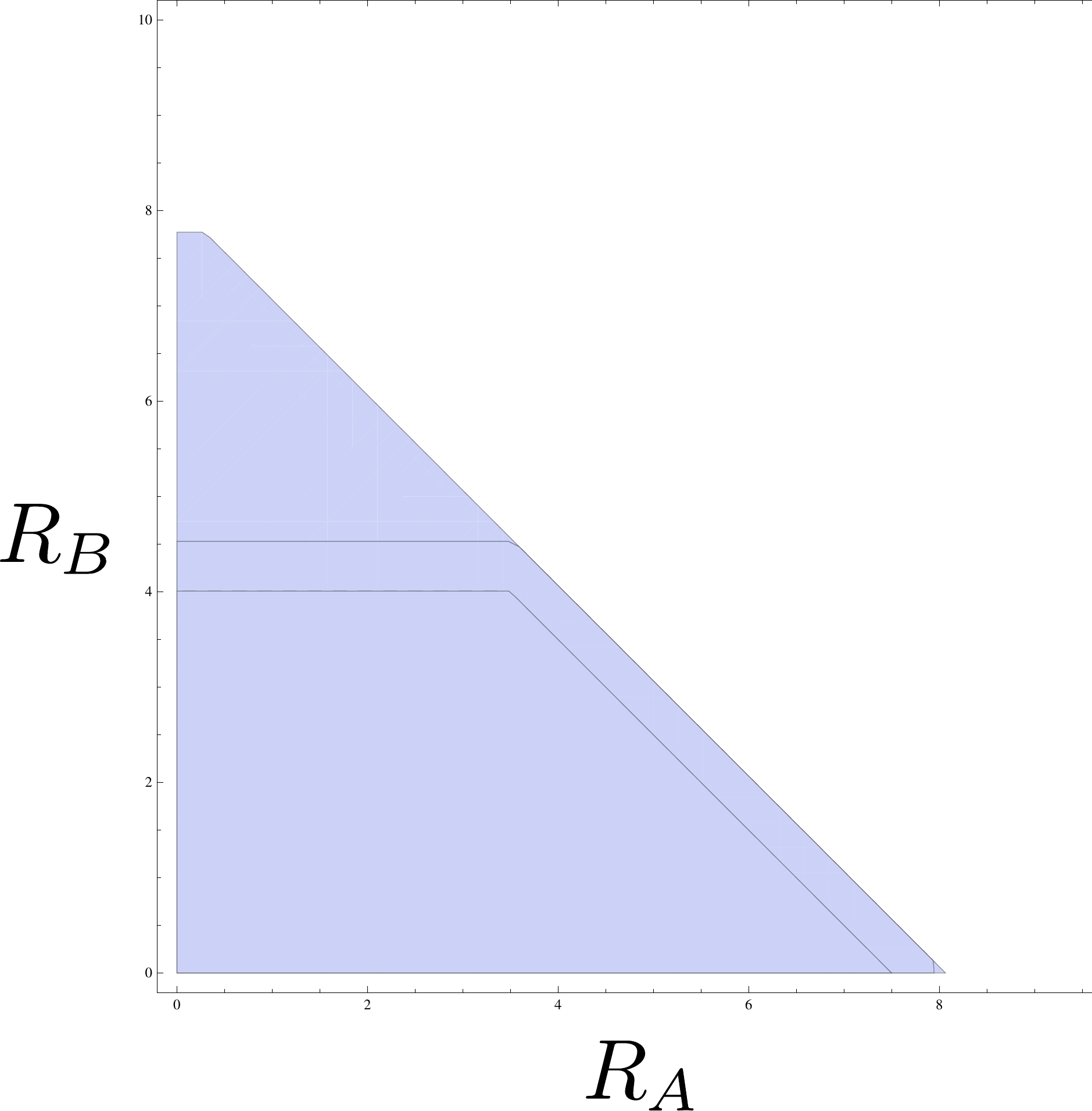}}
\caption{Simple multiple access channel.  (a) Discrete quadrature
  model for the multiple access channel from Fig
  \protect\ref{multiuserchannels}a with $\lambda W_A \geq
  (1-\lambda)W_B$.  Discretizing the input quadratures as usual gives
  $\log \sigma_{P_A} \sigma_{Q_A}/\Delta P_A\Delta Q_A$ for sender $A$
  and $\log \sigma_{P_B} \sigma_{Q_B}/\Delta P_B\Delta Q_B$ for sender
  $B$.  The channel maps input quadratures $P_A,Q_A$ and $P_B,Q_B$ to
  outputs $\sqrt{\lambda} P_A,Q_A + \sqrt{1-\lambda} P_B,Q_B$.  The
  highest order bits from $A$ remain distinguishable at the output
  regardless of what $B$ sends.  The lower order bits are confusable;
  one of the senders can get information through to $C$ reliably so
  long as the other sender holds the corresponding bits fixed.  (b)
  The classical-capacity rate region for the multiple access channel.  
  The outer blue
  region is the upper bound from \protect\cite{ShapiroYen}.  The black
  lines indicate their achievable rate region with coherent states and
  the achievable region for our model with 
  $\Delta P_A=\Delta Q_A$, $\Delta P_B=\Delta Q_B$.  Our region is lower by just a fraction of a bit.
\label{multipleaccessfigure}
}
\end{figure}
\begin{figure}
\centering
\subfloat[]{\includegraphics[height=1.5in]{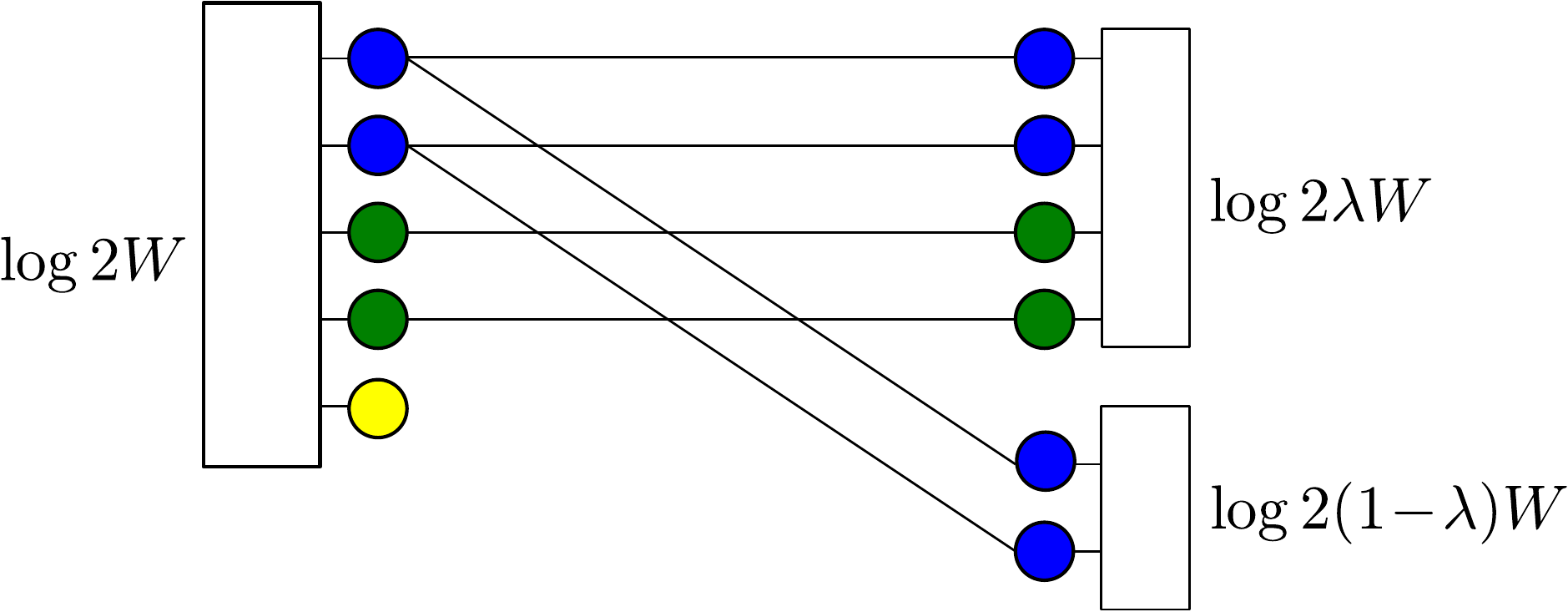}} 
\subfloat[]{\includegraphics[height=1.5in]{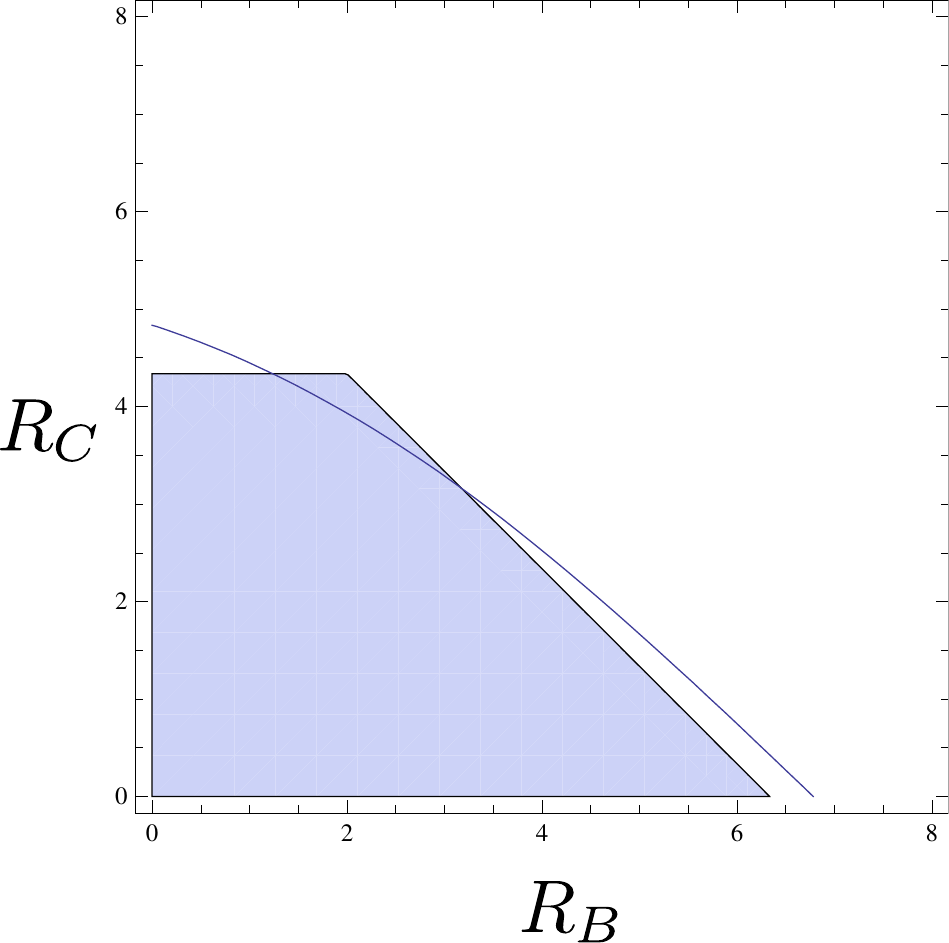}}
\caption{Broadcast channel.  (a) The discrete quadrature model for the
broadcast channel of \protect\ref{multiuserchannels}b.  We have taken
$\Delta P =\Delta Q = 1/\sqrt{2}$ which achieves optimality.  The
rate region is then given by $R_c \leq \log{2 (1-\lambda)W}$ and 
$R_B+R_C \leq \log{2 \lambda W}$.  (b) The rate region is plotted for 
$W=50.5$, $\lambda=.8$ and compared to the lower bound from \protect\cite{GSE} 
which they conjecture to be the actual capacity.
\label{broadcastfigure}
}
\end{figure}

A simple multiple access channel is shown in Fig. \ref{multiuserchannels}a.  
Two senders $A$ and $B$ try to transmit information to a reciever $C$.  
The performance of such a channel is described by a \emph{rate region}
rather than a single capacity.  The discrete quadrature model of this channel
is shown in Fig. \ref{multipleaccessfigure}a and the predicted rate region
is compared to existing bounds in Fig. \ref{multipleaccessfigure}b.
Similarly, the DQ model and rate region for a simple broadcast channel, where one
sender, $A$, tries to send information to two receivers $B$ and $C$, is shown in
Fig. \ref{broadcastfigure}.

\section{Discussion}

Our approach allows us to solve a number of vexing questions that are
intractable in the orthodox model.  For example, we can exactly solve
models for single-user communication with thermal noise, as well as
multi-user networks including broadcast and multiple-access channels.
Perhaps, though, our model is too crude to capture the relevant
behavior of quantum communication systems.  We argue, to the contrary,
that we capture the relevant physics by comparing the few solvable
gaussian examples to our predictions where we find good agreement to
within one or two bits.  Furthermore, for examples where only lower
bounds are available, we find good agreement with these, thus
predicting that typically known lower bounds are equal, or at least
close, to the ultimate capacities.

Our model also explains some previously known but counter-intuitive
facts, rendering them almost obvious.  For instance, we can explain
why while classical capacity rises without bound as power increases,
the quantum capacity saturates: Increasing power enhances transmission
to both receiver and eavesdropper in equal measure.  Put simply, if
you're trying to transmit privately, shouting your secrets doesn't
help.  We have explained the results of \cite{WHG12}, that
time-sharing is not optimal for the classical-quantum tradeoff when
trying to simultaneously transmit some of each type of information.

Finally, we can make some predictions.  Within our deterministic
model, 
entanglement between channel uses and other quantum tricks don't
appear to be useful.  In particular, (1) Privacy and coherence are
equivalent in our model.  Therefore we predict the private and quantum
capacities will always be nearly equal in gaussian channels, even
though they can be very different in general \cite{HHHO2005}.  (2)
Two-way communication doesn't help much for gaussian channels, again
counter to the general case \cite{BDSW96}.  (3) Gaussian channels have
a single-letter capacity forumla to within a small number of bits
(compare to \cite{SY08,SS09}).  (4) Capacities of gaussian channels are
nearly additive (unlike the extreme superadditivity in
\cite{SY08,SS09}).

The ``nearly'' can, however, obscure some interesting effects.  We
know, for example, that gaussian channels can display superactivation,
that is there exist pairs of gaussian channels each with zero capacity
that can neverthelesss be used together to achieve positive capacity
\cite{SSY2011}.  The resolution is that the resulting capacities are
very small (the rate achieved with the joint channel in \cite{SSY2011}
is only 0.06 bits).  Our predictions are only meant to be with a few
bits of the ``correct'' value so there is no contradiction, and our
results should be asymptotically correct at high SNR.

\bibliography{DQ}

\end{document}